\documentclass[a4paper,twocolumn,amsfonts,notitlepage,superscriptaddress,nofootinbib]{revtex4-1}
\pdfoutput=1
\usepackage[T1]{fontenc}
\usepackage[utf8]{inputenc}
\usepackage{amsmath,amssymb,amsfonts}
\usepackage{mathrsfs}
\usepackage{bbm}
\usepackage{slashed}
\usepackage[dvipdf,dvips,dvipdfmx]{graphicx}
\usepackage{verbatim}

\usepackage[english]{babel}
\usepackage{mathbbol}
\usepackage[colorlinks,citecolor=blue,linkcolor=blue,urlcolor=blue]{hyperref}

\def\be{\begin{equation}}
\def\ee{\end{equation}}
\def\ba{\arraycolsep .1em \begin{eqnarray}}
\def\ea{\end{eqnarray}}

\def\apjl{ApJ}%
\def\apjs{ApJS}%
\def\ao{Appl.\ Opt.}%
\def\aap{A\&A}%

\begin{document}

\title{Non-perturbative unitarity and fictitious ghosts in quantum gravity}

\author{Alessia Platania}
\email{a.platania@thphys.uni-heidelberg.de}
\affiliation{Institut für Theoretische Physik, Universität Heidelberg, Philosophenweg 16, 69120 Heidelberg, Germany}
\author{Christof Wetterich}
\email{c.wetterich@thphys.uni-heidelberg.de}
\affiliation{Institut für Theoretische Physik, Universität Heidelberg, Philosophenweg 16, 69120 Heidelberg, Germany}

\begin{abstract}
{We discuss aspects of non-perturbative unitarity in quantum field theory. The additional ghost degrees of freedom arising in ``truncations'' of an effective action at a finite order in derivatives could be fictitious degrees of freedom. Their contributions to the fully-dressed propagator -- the residues of the corresponding ghost-like poles -- vanish once all operators compatible with the symmetry of the theory are included in the effective action. These ``fake ghosts'' do not indicate a violation of unitarity.}
\end{abstract}

\maketitle

\section{Introduction}

A  consistent and fundamental quantum theory of gravity has to be renormalizable and unitary. On the one hand, the Einstein-Hilbert action is perturbatively non-renormalizable, but unitary. On the other hand, the inclusion of  terms with four derivatives in the gravitational action makes the theory renormalizable, but introduces a spin-2 ghost spoiling the perturbative unitarity of the theory~\cite{Stelle:1977ry}.

It has been shown that some classes of non-local theories of gravity can be unitary \cite{Tomboulis:1997gg,Tomboulis:2015esa}. Specifically, when considering an exponential of entire functions, the propagator does not display any ``extra'' ghost poles: at a tree level, this type of non-local theories are unitary \cite{Tseytlin:1995uq}. However, if these theories are considered to be non-local at a fundamental level, i.e., at the level of the bare theory, then quantum effects could generate infinitely many massive complex poles \cite{Shapiro:2015uxa}, leading to the presence of  acausal effects on microscopic scales~\cite{Tomboulis:1977jk,Han:2004wt,Calmet:2013hia}. 
Depending on the scale of the violation, these acausal effects could still be compatible with observations, thus making non-local gravity a viable approach to construct a renormalizable and unitary theory of quantum gravity. However, microscopic locality is one of the fundamental properties of quantum field theory (QFT). Is it possible to construct a unitary and renormalizable theory of quantum gravity whose fundamental (bare) action is local?

From the point of view of QFT, starting from a local fundamental theory, it is the process of resumming quantum fluctuations (quantum loops) at all scales that generates non-localities at the level of the effective action. Indeed, it has been proposed~\cite{Fradkin:1981vx} that quantum corrections may restore unitarity: the expectation is that interaction can make the spin-2 ghost of Stelle-gravity unstable~\cite{Donoghue:2019fcb}, or even remove it from the spectrum of all possible asymptotic states. In fact, due to fluctuation effects, the inverse dressed propagator is not simply linear in~$q^2$, rather it is a non-local function, involving logarithms for example. If interaction is able to remove the ghost from the Fock space of asymptotic states, then the unitarity of the theory is safe. 

A compelling proposal for a theory of quantum gravity based on local QFT is the asymptotic safety scenario for quantum gravity \cite{Niedermaier:2006wt,WIKInink,Eichhorn:2017egq,Pawlowski:2020qer}. 
According to the asymptotic-safety conjecture \cite{1976W}, a (non-perturbatively) renormalizable QFT of gravity can be constructed based on the existence of a suitable non-trivial fixed point of the renormalization
group (RG) flow. The non-perturbative methods of the functional renormalization group (FRG) (see~\cite{Dupuis:2020fhh,Pawlowski:2020qer} for recent reviews), based on Wilsonian idea of renormalization \cite{Wilson:1973jj}, indicate the existence of a fixed point in four dimensions - the Reuter fixed point. So far this has been seen in various truncations of the exact flow equations (see \cite{Percacci:2017fkn,Reuter:2019byg} and references therein). In the framework of the FRG, the effective action can be derived from the flow of the effective average action $\Gamma_{k}$. The effective action is obtained in the limit of vanishing RG-scale, $k\to0$, as in this limit all quantum fluctuations are integrated out. All scattering amplitudes derived from the effective action at a tree level incorporate the effects of all quantum loops, i.e., they are dressed quantities. Thus, important aspects of unitarity and stability are best discussed on the level of the quantum effective action, since all fluctuations effects are included~\cite{Salam:1978fd}. In practical computations however, truncations of the theory space are employed and the presence of a finite number of higher-derivative operators naturally lead to the generation of several poles in the graviton propagator. 

In this letter we  investigate the question whether the ghost-poles of the graviton propagator observed in truncations to the effective action constitute a real problem for the unitarity of the theory, or are rather artifacts of the truncation. This would give a physical answer to criticisms~\cite{Donoghue:2019clr} that asymptotically safe gravity is not unitary~\cite{Wetterich:2019qzx,Bonanno:2020bil,Draper:2020bop,Draper:2020knh}. It will be shown that the inclusion of quantum effects at all scales is crucial to assess unitarity of QFTs. We will also show with explicit examples that poles appearing in truncations of the effective action for a consistent QFT correspond to \emph{fake} degrees of freedom of the theory: their residues are negative only if a few terms in a derivative expansion are considered, while increasing the truncation order the absolute value of the corresponding residues decreases and vanishes once all operators allowed by symmetry are included in the action. We will formulate criteria for a consistent graviton propagator and show the existence of propagators obeying these criteria.

\section{Non-perturbative aspects of unitarity in QFT and quantum gravity}\label{sect2}

Perturbative expansions in QFT are typically used as a tool to simplify computations. While this approach can work for field theories that are perturbative at all scales, it could give the wrong answer for theories where non-perturbative or all-orders effects are important. 


In the following we point out some of the arguments related to the definition of unitarity and based on perturbation theory which could fail for non-perturbative field theories. We also highlight some subtle details and ambiguities that render the issue of unitarity in quantum gravity even more involved. In particular we discuss (apparent) issues with unitarity that can easily arise when employing the FRG to extract the effective action.

\subsection{Non-perturbative optical theorem}

The optical theorem itself, complemented by the full LSZ expansion of the $S$-matrix, do not rely on any perturbative expansion. The fully non-perturbative optical theorem can be represented diagrammatically
as follows
\noindent \begin{center}
\includegraphics[scale=0.37]{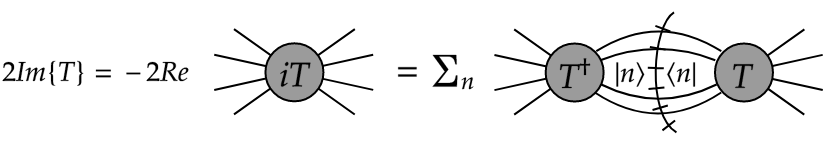}
\par\end{center}
where $T$ is the transfer matrix. The sum in the right-hand-side runs over all possible intermediate states belonging to the space of states of the full (possibly non-perturbative) interacting theory. If there are no negative-norm states in the full theory, the space of asymptotic states is a Fock space and the sum over projectors in the right-hand-side defines the identity in the corresponding Fock space. 

In the standard perturbative approach, the optical theorem  (which follows from the condition that the S-matrix is unitary,  $S^{\dagger}S=\mathbb{1}$) is translated into an infinite set of equalities, and unitarity has to be satisfied at each order in perturbation theory. However, as we will see, if a perturbative expansion breaks down, this would immediately lead to a(n apparent) violation of unitarity. In particular, a theory could violate unitarity at a perturbative level, while being non-perturbatively unitary.

\subsection{Asymptotic states and vacuum}

From a perturbative point of view, asymptotic states are constructed as free-particles states, and defined as excitations over the free-vacuum $|0\rangle$ of the (non-interacting) theory. The in- and out-states must thus be well-separated at asymptotic times, such that interaction can be neglected: in this limit the Heisenberg fields are assumed to become free fields. When the
particles approach each others they start interacting and this interaction
is governed by the full Hamiltonian $H=H_{0}+\lambda H_{int}$, with $\lambda\ll1$ to guarantee that interaction is just a small correction to the free Hamiltonian (this is equivalent to say that the couplings appearing in $H_{int}$ are small). When in a theory interaction or self-interaction is always present (e.g., when the bare theory is not free), asymptotic states and initial propagation should be defined
using the fully-interacting theory \cite{Kulish:1970ut,Horan:1999ba,Dybalski:2004cj}. This means that in/out states
should be eigenstates (stable particles or bound states) of the fully non-perturbative Hamiltonian and should be defined as excitations over the (non-perturbative) vacuum of the full theory $|\Omega\rangle$.  

In the case of gravity, it is not even obvious that the Minkowski spacetime is the true vacuum of the theory. Even in the simple case of quadratic gravity, at least in its conformally-reduced version, the dominant configuration in the gravitational path integral could correspond to a complicated ``kinetic condensate'' \cite{Bonanno:2013dja,Bonanno:2019pmm}, rather than a ``simple'' flat spacetime. While this result depends on the structure of the full theory, it is important to keep in mind that a proper definition of an $S$-matrix requires the knowledge of the asymptotic states~\cite{Wiesendanger:2012is} about the true vacuum of the theory, and that the latter might be non-trivial in the case of gravity. 

\subsection{Effective actions, scattering amplitudes, non-perturbative unitarity and truncations}

The quantum effective action $\Gamma_0$ encodes the effects of all quantum loops and is the generator of 1PI Green functions. Thus, all scattering amplitudes (propagators and vertexes) computed at a tree level using $\Gamma_0$ are fully-dressed, i.e., they already contain the effects of all quantum loops. These are given by the functional derivatives of the effective action:
\begin{equation}
\begin{aligned}
\langle f|S|i\rangle\propto&\langle\Omega|T\left\{ \phi(x_{1})\dots\phi(x_{n})\right\} |\Omega\rangle_{(c)}=\\
&\left[\frac{\delta^{n}\Gamma_0[\phi]}{\delta\phi(x_{1})\dots\delta\phi(x_{n})}\right]_{\phi=0}\,,
\end{aligned}
\end{equation}
where $\phi(x_{i})$ are fully interacting quantum fields and $|\Omega\rangle$ is the vacuum of the fully interacting theory. As the quantum effective action includes all (perturbative or non-perturbative) effects of quantum loops at all momentum scales, it can be used to verify unitarity in both perturbative and (strongly or weakly) non-perturbative QFTs. 

The effective action $\Gamma_0$ can be obtained either by solving the functional integral of a theory or via the FRG equation \cite{Wetterich:1992yh}
\begin{equation}\label{floweq}
k\partial_{k}\Gamma_{k}=\frac{1}{2}\mathrm{{STr}}\left\{ \left(\Gamma_{k}^{(2)}+\mathcal{R}_{k}\right)^{-1}k\partial_{k}\mathcal{R}_{k}\right\} \,\,.
\end{equation}
Here $\mathcal{R}_{k}$ is a regulator function and $\Gamma_{k}^{(2)}$ denotes the second functional derivative of the effective average action~$\Gamma_{k}$. The latter is a RG-scale-dependent effective action, which results from the integration of fluctuating modes with momenta $p\in(k,\infty)$. The quantum effective action is thus obtained as the limit $k\to0$ of $\Gamma_k$ and is expected to be non-local (even when starting from a local bare or microscopic action, $S_c=\Gamma_\infty$) due to the integration of quantum fluctuations at all scales.

The FRG turned out to be a powerful tool to study the (non-perturbative) renormalizability of field theories and explore their implications for infrared physics. Nevertheless, one of the 
drawbacks of the FRG is the practical necessity to ``truncate'' the theory space, i.e., to use a truncated (derivative or vertex) expansion of $\Gamma_k$, in order to  solve Eq.~\eqref{floweq} and derive $\Gamma_0$. 
While in the case of field theories which are perturbative at all scales it might be sufficient to consider only operators with positive or zero mass dimension, in general $\Gamma_k$ should contain \emph{all} possible operators allowed by symmetry. In the case of gravity, this means that $\Gamma_k$ should contain all operators compatible with diffeomorphism invariance. While the truncated-FRG computations still allow to explore the existence of fixed points of the RG flow, it is clear that once a truncated derivative expansion for the effective action $\Gamma_k$ is employed, the propagator will automatically display additional (ghost or tachyon or tachyonic ghost) poles, which could just be an artifact of the truncation, rather than a problem for the theory. In particular, this could be the case for the ghost of Stelle gravity.

\section{Non-perturbative unitarity in one-loop QED and Lee-Wick QED}\label{sect3}

In the following we use QED as a working example to show how fictitious ghost poles can appear in an artificially-truncated version of the theory. 

As effective actions are typically non-local, we can assume the
quadratic part of the QED-effective action to take the form
\begin{equation}
\Gamma_0^{QED}[A_{\mu}]=-\frac{1}{4}\int d^{4}x\left\{ F_{\mu\nu}P(\square)F^{\mu\nu}\right\}\,. \label{eq:toyQED}
\end{equation}
The latter has to be complemented with a gauge fixing term
\begin{equation}
S_{\text{gf}}=-\frac{1}{2\xi}\int d^{4}x\left\{ \partial_{\mu}A^{\mu}Q(\square)\partial_{\nu}A^{\nu}\right\} \,,
\end{equation}
and the corresponding propagator reads
\begin{equation}
\Delta_{\alpha\beta}(q^{2})=\frac{i}{q^{2}P(q^{2})}\left\{ \eta_{\alpha\beta}-\left(1-\xi\frac{P(q^{2})}{Q(q^{2})}\right)\frac{q_{\alpha}q_{\beta}}{q^{2}}\right\} \,.\label{eq:propagator}
\end{equation}
In what follows we will fix $\xi=0$. We now need to specify the form of $P(q^2)$. Following~\cite{LandauD,Donoghue:2015nba,Donoghue:2015xla}, at one loop this function reads
\begin{equation}
P(q^{2})=1-\frac{\alpha}{3\pi}\log\left(\frac{-q^{2}+m_{th}^{2}}{m_{th}^{2}}\right)\,,\label{eq:Pfunction}
\end{equation}
where $\alpha$ is the fine structure constant and $m_{th}$ is a threshold mass, $m_{th}^2=4m^2$, with $m$ being the mass of the degree of freedom integrated out to obtain the one-loop effective action, typically the electron mass. Due to the presence of the logarithm, there is a branch cut singularity, corresponding to the production of particles. The scalar part of the propagator $D(q^2)=q^{-2}P^{-1}(q^2)$, with the function $P(q^2)$ given in Eq.~\eqref{eq:Pfunction}, has no poles in the regime where the theory is valid, i.e., for momenta $q^2\gtrsim q^2_L$, with $q_L^2\sim -10^{560}m_{th}^2$ being the Landau pole.

In what follows, we will use~$\eqref{eq:toyQED}$ as a \emph{toy model} for the quadratic part of the full QED effective action, with $P(q^2)$ given in Eq.~\eqref{eq:Pfunction}. 


Although effective actions are generally non-local, when expanding
the effective action to get a low-energy effective description of
the theory, the action can be expressed as a series of local terms.
This corresponds to an energy, or derivative expansion, in which $q^{2}$ is small as compared to the mass of the degree of freedom that has been integrated out. 

\begin{figure}[t!]
\hspace{-0.4cm}\includegraphics[scale=0.46]{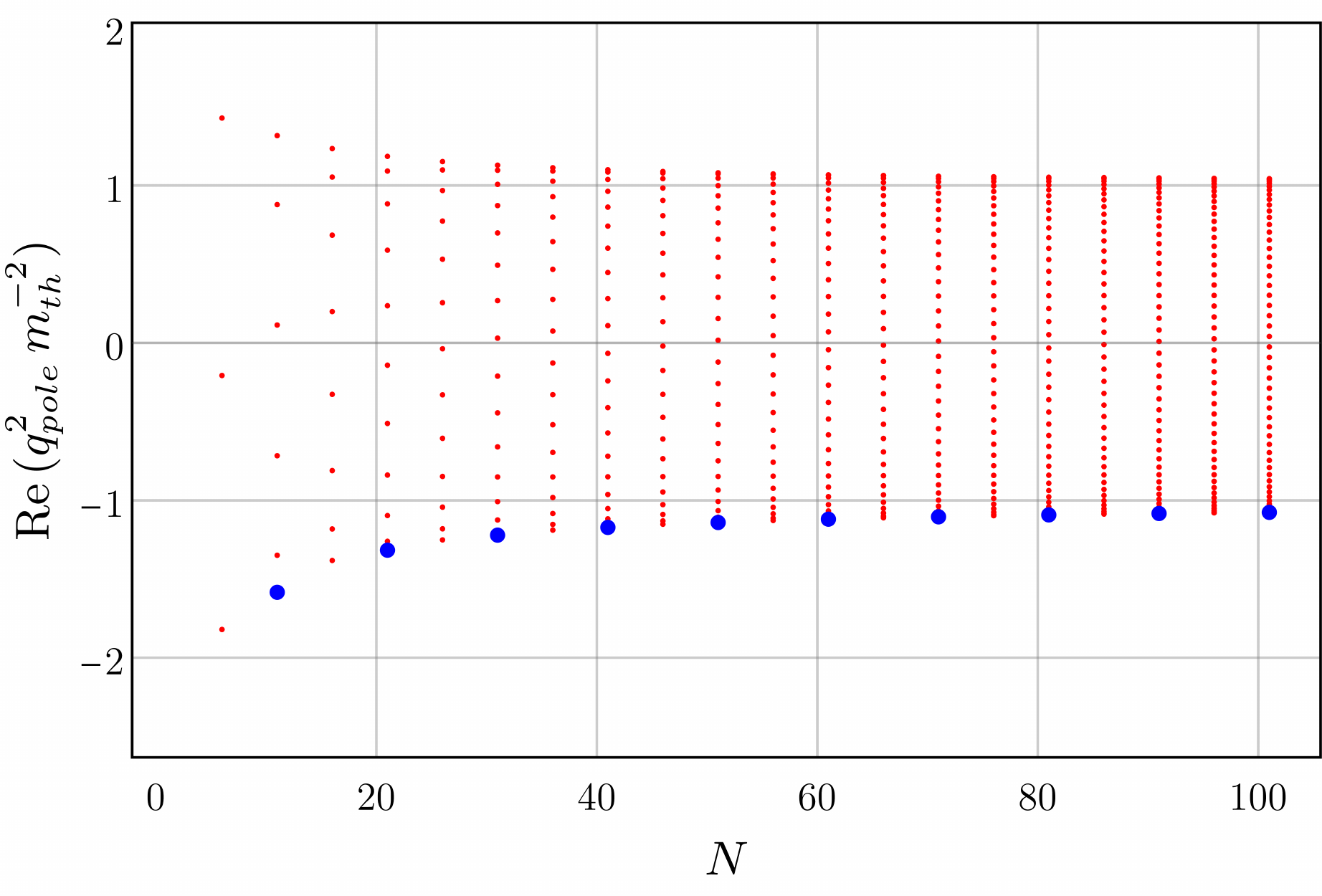}
\caption{Real part of the poles of the truncated one-loop QED propagator as function of the truncation order $N$. The truncated propagator has several complex conjugate poles (red dots) and one tachyonic ghost-pole (blue dots) whose location approaches $q^2=-m_{th}^2$ for large truncation orders~$N$. \label{PolesExpandedPropy1}}
\end{figure}

Defining $z\equiv q^{2}/m_{th}^{2}$, the expansion about $z=0$ of the function $P(z)$ to the truncation order $N$ reads
\begin{equation}
P_{N}(z)=1+\frac{\alpha}{3\pi}\sum_{n=1}^{N}\frac{z^{n}}{n}\,.\label{eq:expansion}
\end{equation}
The first term of this expansion reproduces classical electrodynamics. 
Although the fully-dressed propagator~\eqref{eq:propagator}
with $P(q^{2})$ given by~\eqref{eq:Pfunction} has a unique
pole at $q^2=0$, the function $P_{N}(z)$
can show additional real and complex-conjugate zeros. In the case at hand, when $N$ is odd the function $P_{N}(z)$ shows to have
a zero at $z\simeq-1$, i.e. $q^{2}=-m_{th}^{2}$, corresponding to
a \emph{stable tachyonic ghost} and entailing an apparent violation of unitarity. In addition, the function $P_{N}(z)$ has several complex-conjugate poles, as shown in Fig.~\ref{PolesExpandedPropy1}. The fact that the ghost is also a tachyon and the fact that it appears
only for $N$ odd depends on the numeric factors in the effective
action. The fact that it is a ghost, i.e., that it comes with negative residue, comes instead from generic properties of polynomials.

The presence of the tachyonic ghost leads to an apparent violation of unitarity: while this ghost does not appear in the full  theory~\eqref{eq:toyQED}, it does if one performs a perturbative expansion of the effective action. As we started from a toy model for the full effective action and we performed a derivative expansion afterwards, it is easy to realize that the tachyonic ghost at $z\simeq-1$ is a truncation artifact. In general however the form of the fully-quantum effective action is not known a priori. It is thereby important to understand in detail the origin of this additional degree of freedom and to come up with conditions to understand \emph{a priori}, namely, without knowing the form of the fully-quantum effective action~$\Gamma_0$, whether a pole is a genuine or {fake} degree of freedom of the theory, i.e., a pole appearing in the full theory or a truncation artifact, respectively.

\begin{figure}[t!]
\hspace{-0.2cm}\includegraphics[scale=0.55]{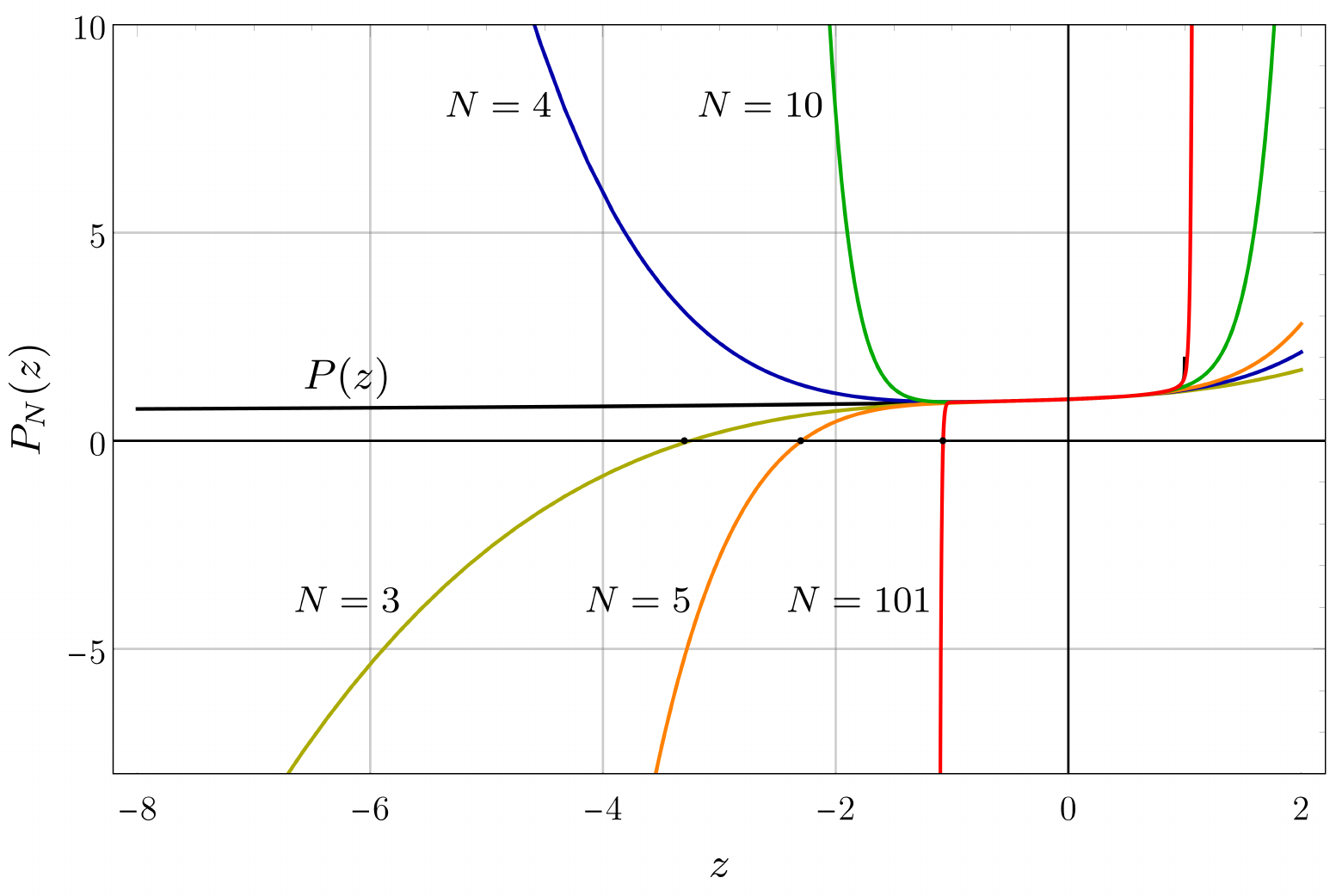}
\caption{Function $P_N(z)$ for different truncation orders $N$. Increasing the truncation order $N$, the function $P_N(z)$ gives better and better approximation to the untruncated function $P(z)$. For $z<0$, the truncated function $P_N(z)$ alternates positive and negative divergences: a zero is generated whenever $P_N(z)$ diverges negatively, i.e., for $N$ odd. These zeros, corresponding to poles of the truncated  propagator are only fictitious: they are not present in the full theory. The position of the zero quickly approaches the boundary of the domain of convergence of the logarithm, localed at $z=-1$.  \label{negposdiv}}
\end{figure}

In the case at hand, the appearance of the additional ghost is due to the convergence properties of the logarithm
at $z=-1$ and, in particular, to the fact that the logarithm has a finite radius of convergence, $|z|<1$. We can understand how the fictitious pole is generated by visualizing the behavior of~$P_{N}(z)$ for increasing values of~$N$. This is shown in Fig.~\ref{negposdiv}.
As we see from the figure, when~$N$ is even $P_{N}(z)$
diverges positively. On the other hand, when~$N$
 is odd, $P_{N}(z)$ diverges negatively and crosses
the $z$-axis, thus generating a pole in the propagator. In particular, as the truncation order~$N$ is increased, the position of the pole approaches the boundary of domain of convergence of the logarithm\footnote{This argument is not restricted to a logarithmic effective action. If the effective action contains a $P(q^2)$ with one or more branch cut singularities, then there will be fictitious poles approaching the boundaries of the domain of convergence of the function~$P$~\cite{PlataniaWetterich}. If instead the integration of quantum fluctuations in the path integral leads to an effective action defined by an entire non-local function, there is no branch-cut singularity and the fake poles slowly move to infinity~\cite{PlataniaWetterich}.}. 
In the limit $N\to\infty$, $P_{N}(z=-1)$ converges
to a finite value and therefore in this limit (equivalent
to say, when the action is not truncated) the fictitious pole disappears. 

\begin{figure}[t!]
\hspace{-0.2cm}\includegraphics[scale=0.43]{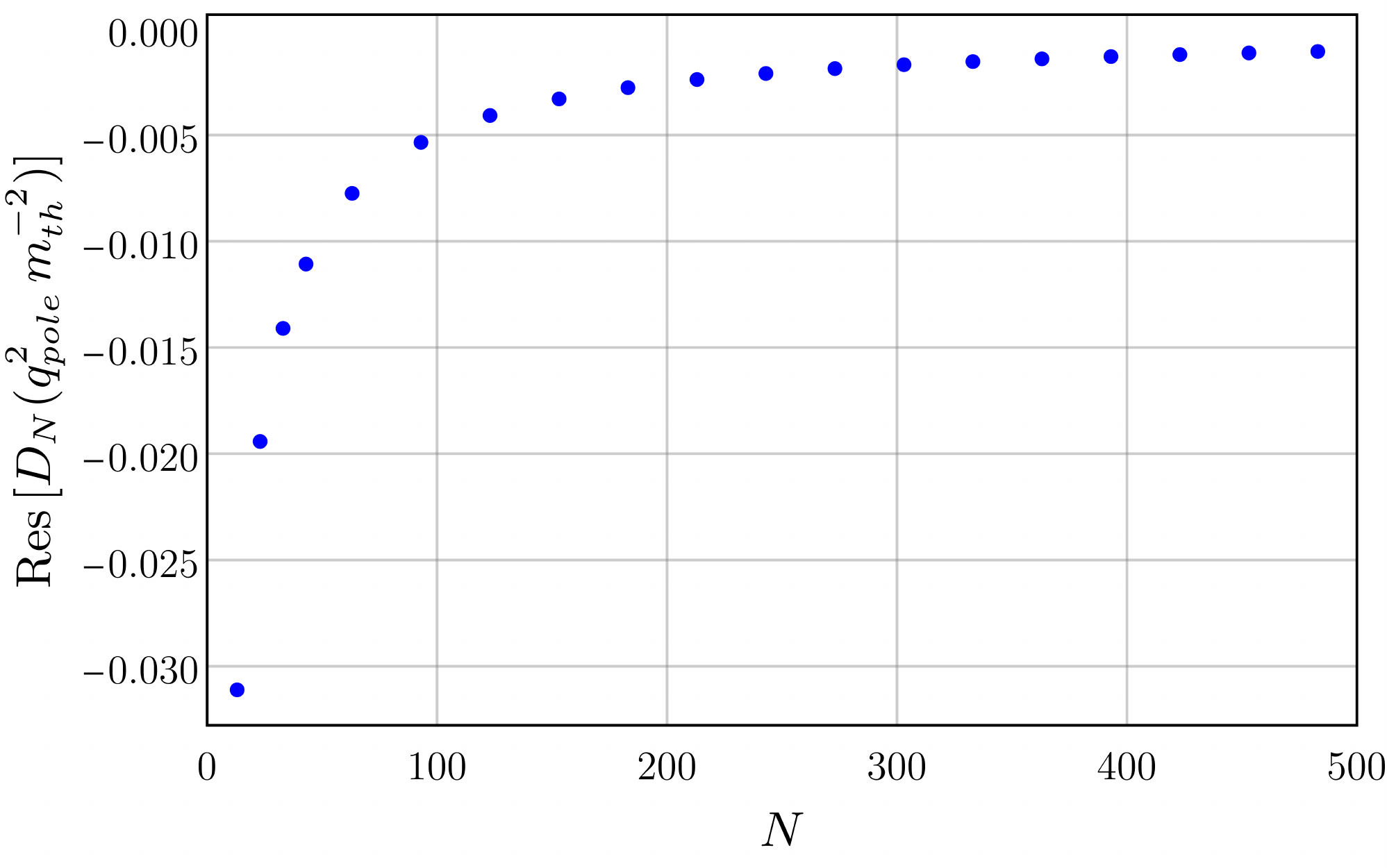}
\caption{Residue of the truncated propagator evaluated at the fictitious tachyonic ghost pole, as function of $N$. As the truncation order is increased the residue approaches zero, thus  making the corresponding fake degree of freedom ``confined''. \label{residuezero}}
\end{figure}

The full (exact) form of the effective action is not known a priori and, especially within the framework of the FRG, it is often necessary to work within a truncation. It is then a key question, if we one can decide a priori whether a pole corresponds to a genuine or fake degree of freedom of the theory. A possible answer lies in its residue. In fact, it turns out that the residue of the propagator at the fake pole decreases by increasing the truncation order $N$ and vanishes in the limit $N\to\infty$, as shown in Fig.~\ref{residuezero}. The reason is that in this limit there is no well-defined particle associated with this pole, and therefore, it cannot give any contribution to the fully-dressed propagator. Interestingly, this is the same mechanism realized in the quasi-particle approach to a Bose-Einstein condensate with impurities~\cite{ChristensenQuasiparticle}, the impurity being the equivalent of the fake ghost pole in QFT.

\begin{figure}[t!]
\hspace{-0.4cm}\includegraphics[scale=0.46]{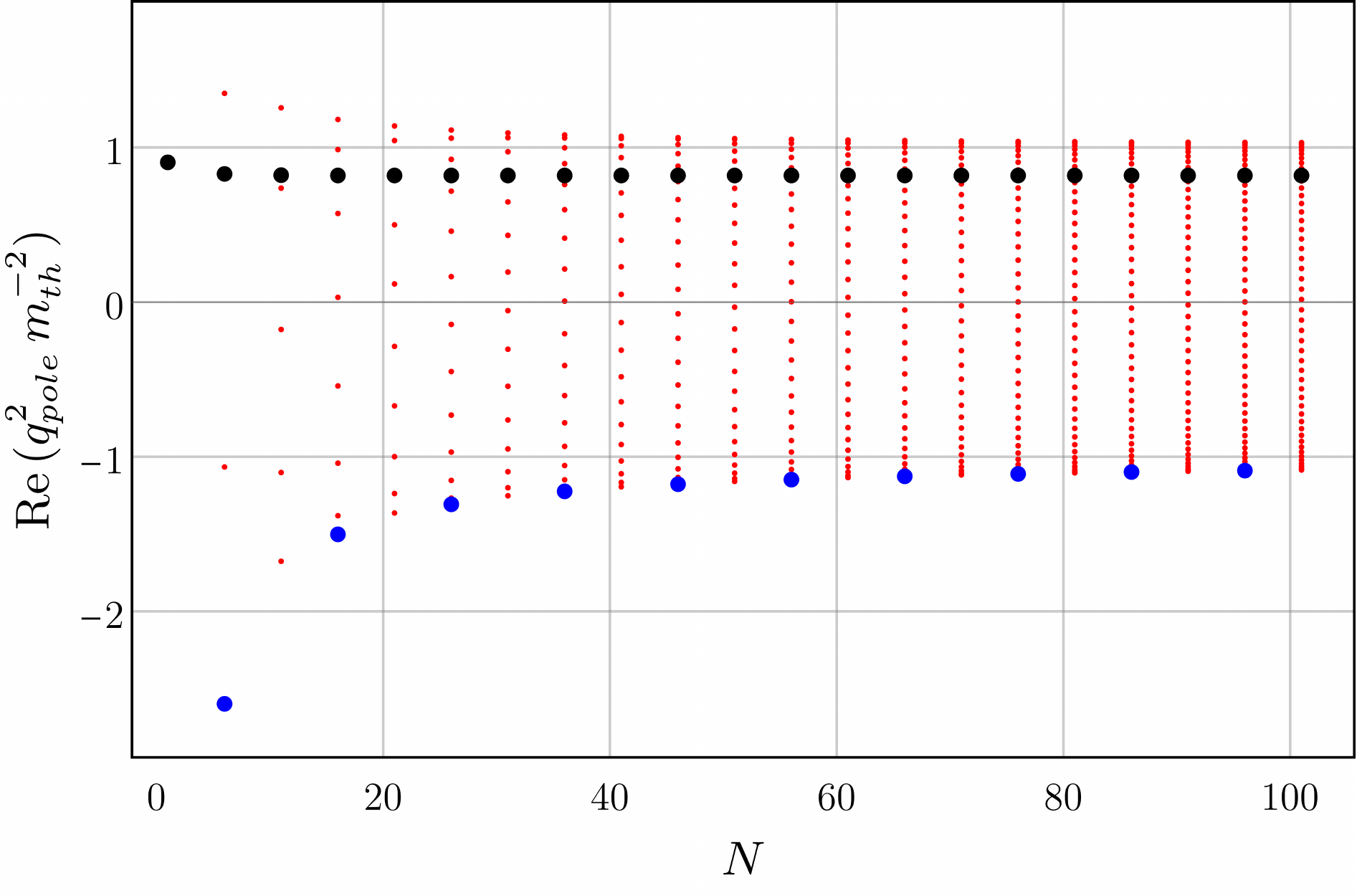}
\caption{Real part of the poles of the Lee-Wick model with opposite sign of the coupling. In this case, beyond a fake degree of freedom represented by a blue dot, there is also an additional ghost pole (black dot) whose real part lies well within the domain of convergence of the function $P_N(z)$. This indicates that this ghost pole will also be present in the ``full theory'', as it can be explicitly checked using the propagator from the untruncated function~$P(z)$. \label{polesleewicknegative}}
\end{figure}

\begin{figure}[t!]
\hspace{-0.2cm}\includegraphics[scale=0.43]{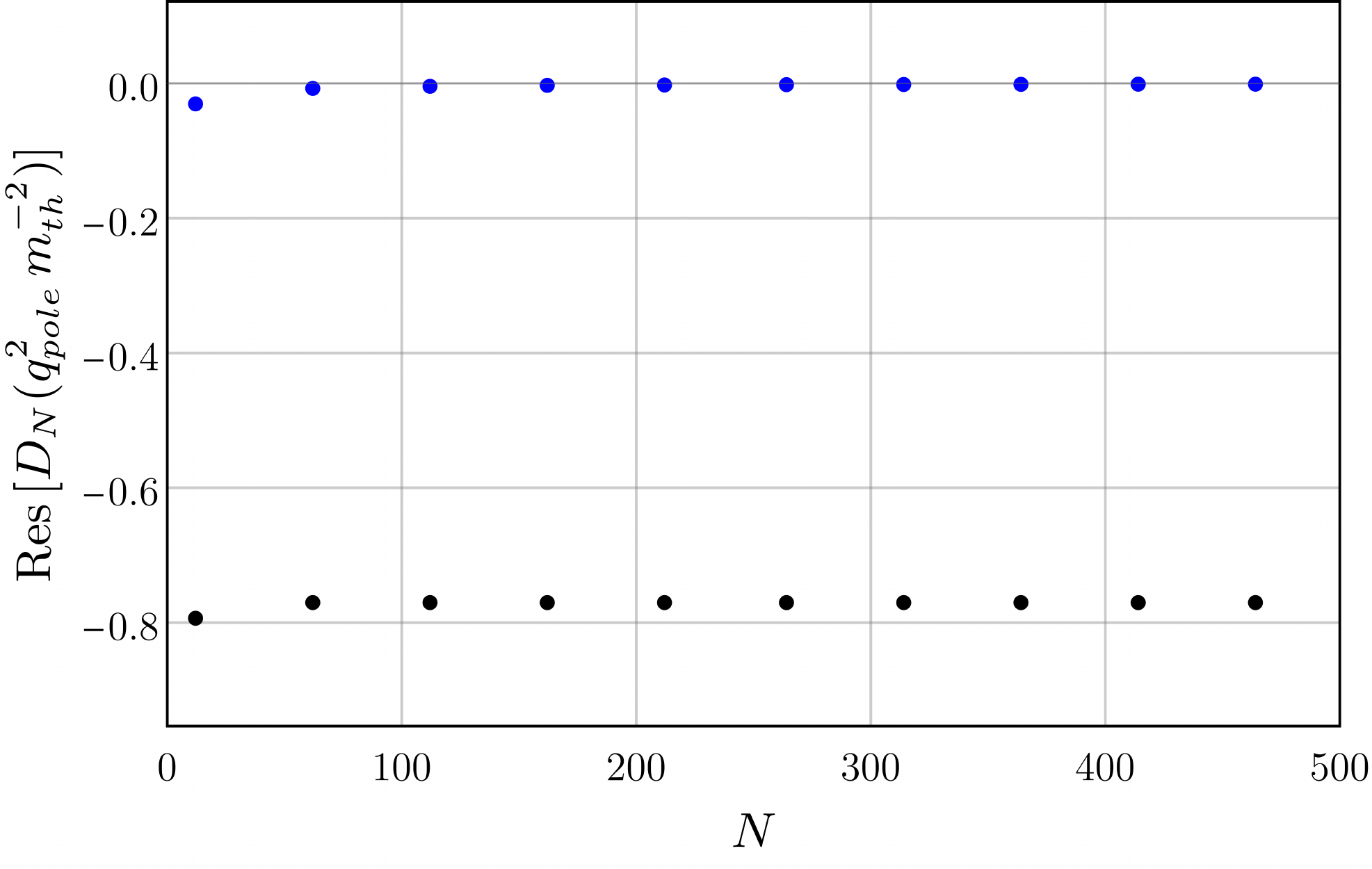}
\caption{Residues of the Lee-Wick propagator at the fake (blue dots) and real (black dots) ghost degrees of freedom, as function of the truncation order. While the residue for the real ghost present in the ``full theory'' is negative and stays negative, the residue of the fake degree of freedom quickly approaches zero as the truncation order is increased. \label{residiuesleewicknegative}}
\end{figure}

As a second instructive example, let us analyze the function
\begin{equation}
P(q^{2})=1+\frac{\alpha}{3\pi}\log\left(\frac{-q^{2}+m_{th}^{2}}{m_{th}^{2}}\right)-\frac{q^{2}}{M^{2}}\,,\label{eq:Pfunction-1}
\end{equation}
which corresponds to a Lee-Wick model for QED, with a coupling whose sign is opposite with respect to the standard one. 
In this case $P(q^{2})$ has a real pole on the principal branch of
the logarithm, corresponding to a stable, massive ghost. An expansion of $P(q^{2})$ about $q^{2}=0$ will generate again a  tachyonic ghost and several complex-conjugate poles, but there will also be a stable, massive ghost for all $N$ (cf. Fig.~\ref{polesleewicknegative}) which lies well within the domain of convergence of the logarithm and does not approach the boundary of its domain of convergence for increasing values of $N$.
This is in fact the ghost appearing in the ``full theory'',
and for this reason the corresponding residue is expected to
stay negative as $N$ is increased. This is indeed the case, as shown in Fig.~\ref{residiuesleewicknegative}: while the residue of the fictitious tachyonic ghost approaches zero for large $N$, the residue of the ghost present in the ``full theory'' quickly stabilizes to a constant negative value.
These results indicate that it might be possible to determine
the nature of the ghosts appearing in truncations of the effective
action, by studying their residues as functions of the truncation order $N$. In particular, this might have important implications for the case of gravity, to understand whether the ghost of Stelle theory is a true ghost or a truncation artifact.

\section{Good propagators}\label{sect4}

Based on the requirements of unitarity, causality, and the possibility of performing an analytic Wick rotation connecting the Euclidean and Lorentzian theories, a fully-dressed propagator should have:
\begin{itemize}
\item No complex poles in the first and third quadrants of the physical sheet of the $q_{0}$-complex plane
\item No essential singularities at infinity
\item Positive-definite spectral density
\end{itemize}
A propagator with these properties can arise from a consistent theory which is valid to infinitely short distances. For a valid QFT of gravity based on asymptotic safety, these criteria have to be obeyed. 

\begin{figure}
\begin{centering}
\hspace{-0.8cm}\includegraphics[scale=0.48]{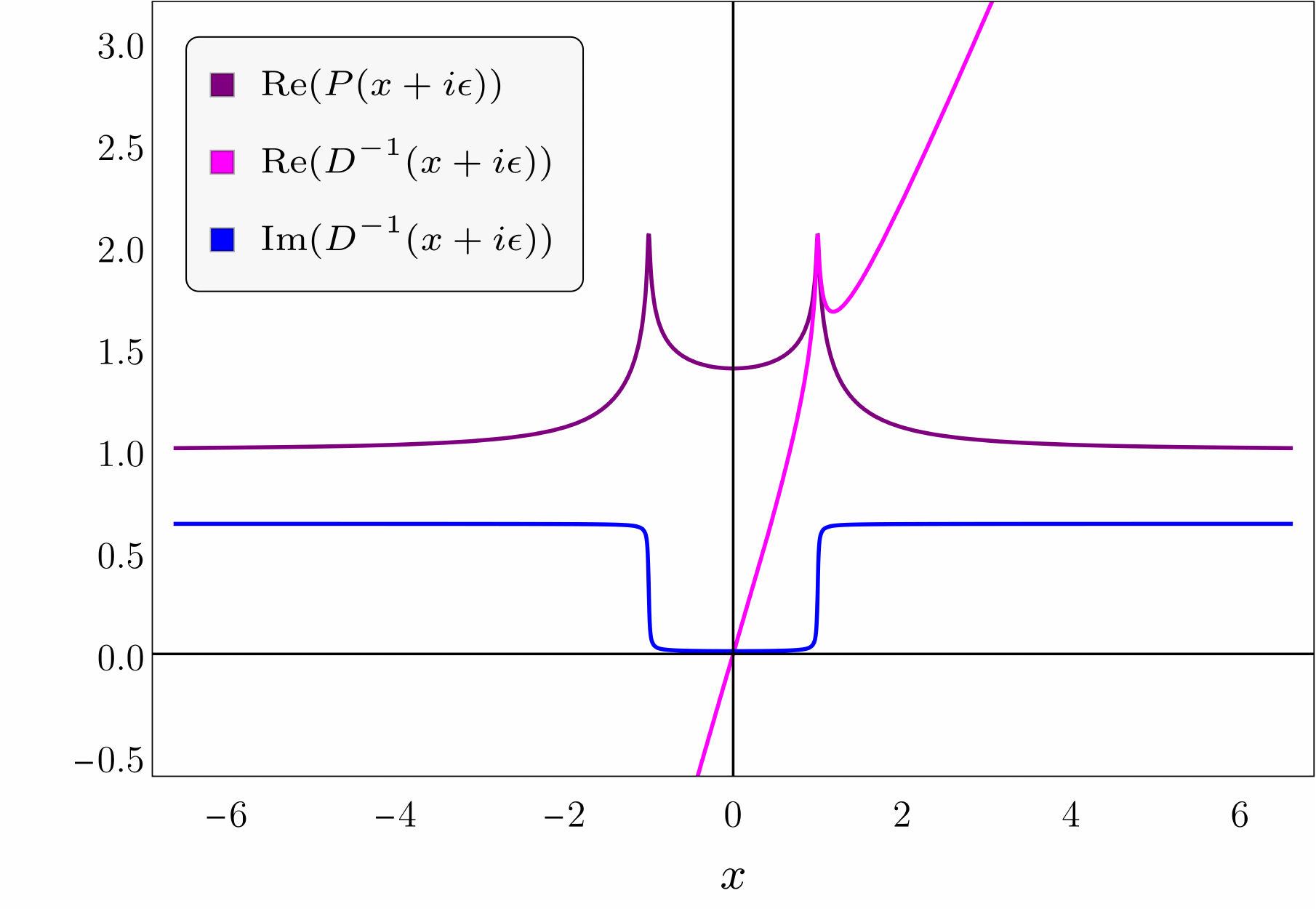}
\end{centering}
\caption{Real and imaginary parts of the function $P(z)$ and of the inverse propagator $D^{-1}(z)$, with $z=x+i\epsilon$. The real part of $P(z)=z^{-2}D^{-1}(z)$
(purple line) is always positive for $\alpha<0$ ($\alpha=-0.2$ in the figure), while the imaginary part of $D^{-1}(z)$ (blue line) is non-zero only along the branch cuts. Accordingly the real part of the inverse propagator $D^{-1}(z)$ (magenta line) has only one pole at $x=0$, corresponding to the massless graviton/photon pole. No additional ghost-like or complex conjugate poles are present.\label{fig:ReImPropy}}
\end{figure}

We next demonstrate by an explicit example that suitable functions $D(q^2)$ exist which obey all criteria. A propagator satisfying all these requirements is\footnote{The vanishing of the propagator at $|x|=1$ and $y=0$ may look somewhat strange, but poses no problem. We do not believe that this feature is essential for the existence of a good propagator. Once one realizes the essential features it seems likely that a large family of good propagators exist. We also note that an expansion of $P(q^2)$ in linear (or any other finite) order in $q^2$ will produce the fake ghost poles discussed previously.}
\begin{equation}
iD(q^{2})=\frac{i}{q^{2}\left(1+\alpha\,\frac{m_{th}^{2}}{q^{2}}\,\,\mathrm{arctanh}\left[-\frac{q^{2}}{m_{th}^{2}}\right])\right)}\,\,,\qquad\alpha<0\label{eq:goodpropy}
\end{equation}
where $m^{2}_{th}$ is a mass scale.
The coupling $\alpha$ must be negative
in order to avoid ghost or complex-conjugate poles. This fact can
be easily seen from the form of the real part of the function $P(z)=z^{-1}D^{-1}(z)$,
with $z=m_{th}^{-2}q^{2}$. It reads
\begin{align}
& \mathrm{Re}(P(z=x+iy))= \nonumber \\
& \quad \,\,1+\frac{\alpha}{4}\frac{x}{x^{2}+y^{2}}\,\log\left(\frac{(1-x)^{2}+y^{2}}{(1+x)^{2}+y^{2}}\right)  \\ 
 & \quad+\frac{\alpha}{2}\frac{y}{x^{2}+y^{2}}\left(\mathrm{arg}(1-x-iy)-\mathrm{arg}(1+x+iy)\right). \nonumber
\end{align}
If $\alpha$ is taken to be negative, then $\mathrm{Re}(P(z=x+iy))\geq1$
$\,\,\forall(x,y)\in\mathbb{R}^{2}$. This implies that for $\alpha<0$
the function $P(z)$ can never be zero, i.e., $D(q^{2})$ cannot have
any pole beyond the massless one (cf. Fig.~\ref{fig:ReImPropy}).
On the other hand, if $\alpha>0$ there might be both ghost-like
poles or complex-conjugate poles. In order to avoid ghosts and maintain unitarity, we thus require the coupling $\alpha$ to be negative.

\begin{figure}[t!]
\hspace{-0.5cm}\includegraphics[scale=0.47]{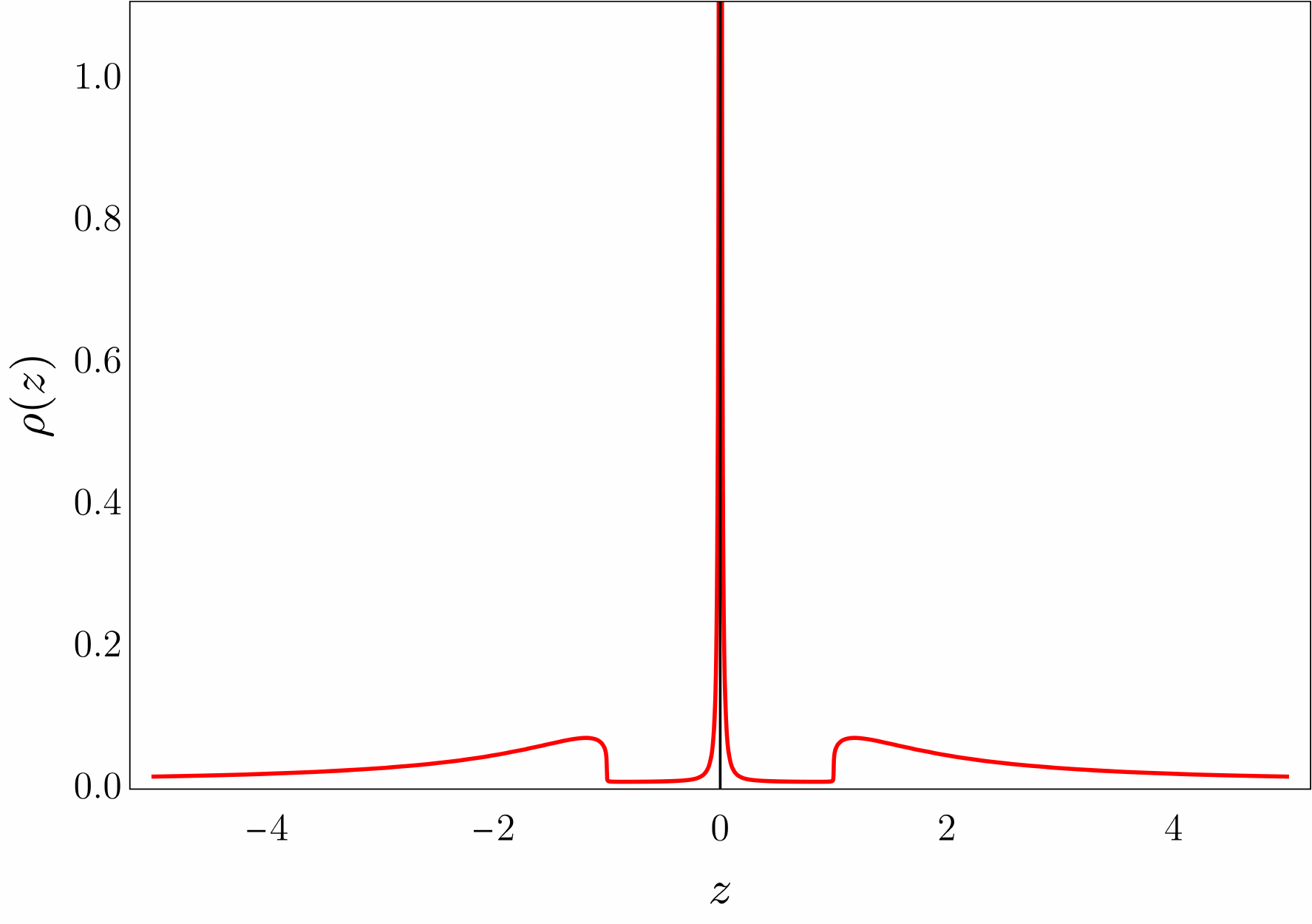}
\caption{Spectral density $\rho(z)$, with $z=q^{2}m_{th}^{-2}$ of the propagator $\eqref{eq:goodpropy}$ for $\alpha=-0.4$. The spectral density is positive definite, indicating that there are no negative-norm states in the theory. The peak at $z=0$ corresponds to the stable massless pole at $q^{2}=0$. For $z>1$, i.e., for $|p^{2}|>m_{th}^{2}$, the spectral density is non-zero due to the branch cuts of the propagator $\eqref{eq:goodpropy}$, describing two disjoint sets of multiparticle states, $m_{th}^{2}$ being the threshold to open the correspondent scattering channels.\label{fig:Spectral-density}}
\end{figure}

The fact that for $\alpha<0$ the real part of
$P(z)$ can never be zero has an important implication: the are no additional stable (ghost) degrees of freedom and the two branch cuts are not associated to a (ghost) resonance as in \cite{Donoghue:2018lmc,Donoghue:2019fcb},
rather to multi-particle states, which are produced for $|p^{2}|>m_{th}^{2}$. In particular, the spectral density $\rho(p^{2})=-\pi^{-1}\mathrm{Im}(D(q^{2}+i\epsilon))$ is positive-definite for $\alpha<0$, as shown in Fig.~\ref{fig:Spectral-density}.

\begin{figure*}[t!]
\begin{centering}
\includegraphics[scale=0.55]{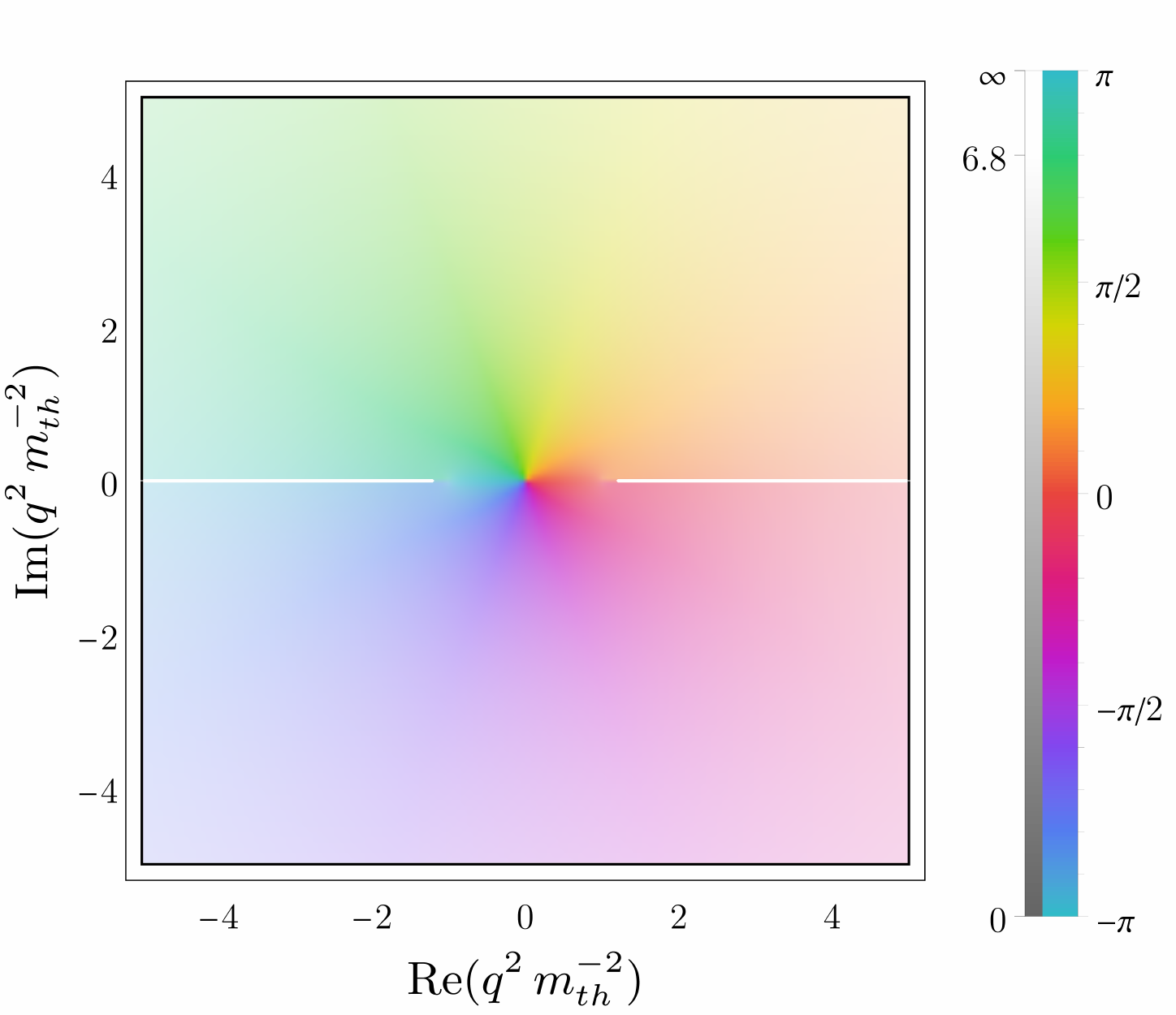}\hspace{0.7cm}\includegraphics[scale=0.55]{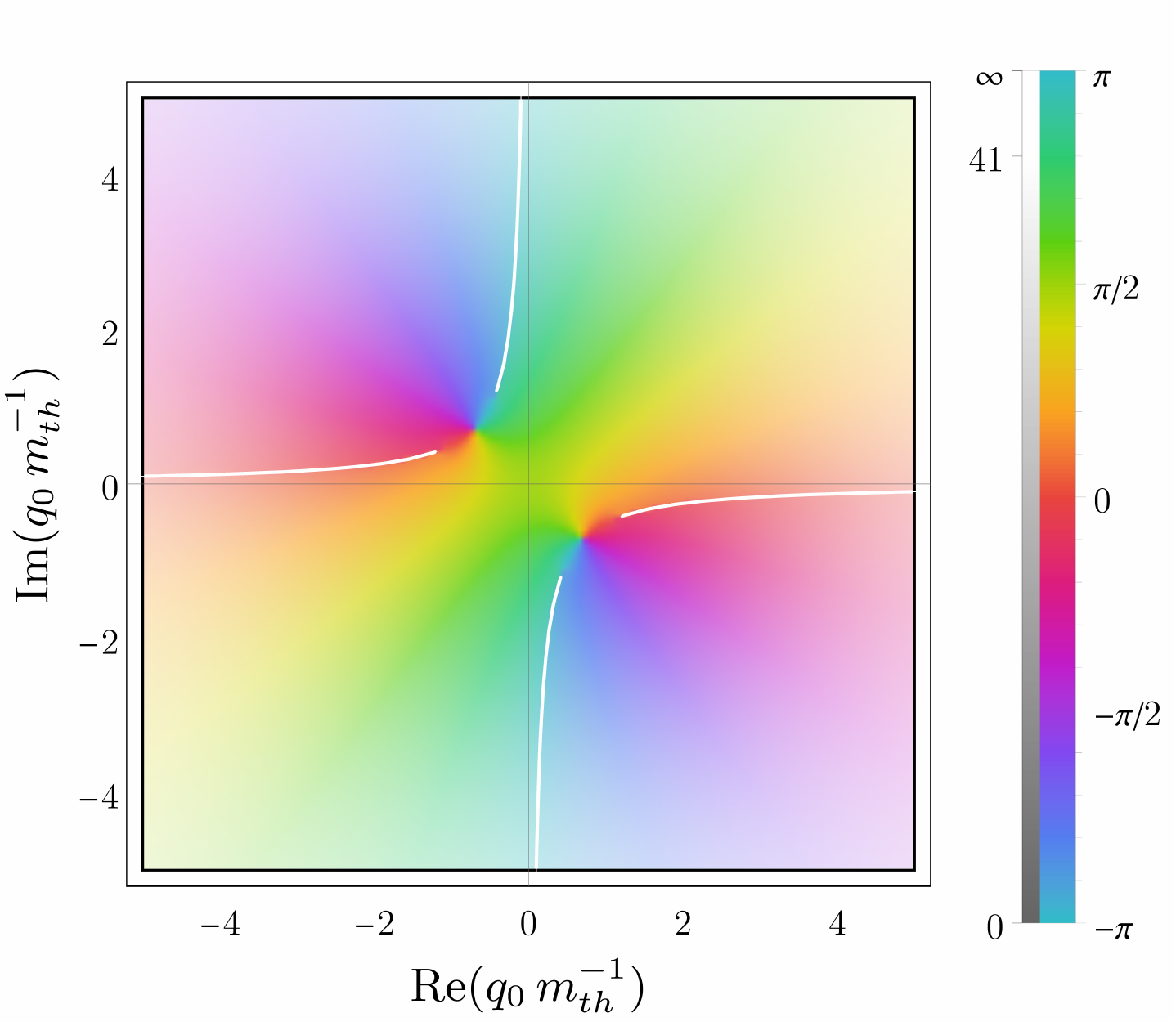}
\end{centering}
\caption{Pole structure of the propagator~\eqref{eq:goodpropy} for $\alpha<0$. In the complex
$q^{2}$-plane (figure on the left panel), there are two branch cuts,
but no additional poles beyond the massless one. The figure on the right panel shows the pole structure of the full propagator $D(q_{0}^{2}-\vec{q}^{2}+i\epsilon)$
in the complex $q_{0}$-plane, with $\epsilon>0$. For $\alpha<0$ there is no obstruction towards performing an analytic Wick rotation from Lorentzian to
Euclidean and vice versa. The theory is thus causal and Wick-rotatable. Both figures have been produced using $\alpha=-0.4$. In the second figure $\epsilon$ has been set to $\epsilon=1$ in order to make
the effect of the Feynmann prescription on the branch cuts visible.\label{fig:Poles-structure}}
\end{figure*}

In our example, the key to avoid stable/unstable,
standard or tachyonic ghosts, is the presence of two symmetric branch
cuts\footnote{The presence of a branch cut at $p^{2}<0$ is not a problem for unitarity
nor for causality: the spectral density can still be positive-definite
(as in our case, cf. Fig.~\ref{fig:Spectral-density}), and the localized
excitations of the multiparticle states at $p^{2}<0$ propagate subluminally
(only the group velocity can be superluminal) \cite{Aharonov:1969vu}.
The theory is thus causal in the sense of QFT, i.e., the commutators/anticommutators of operators at spacelike-separated points are zero \cite{Aharonov:1969vu}. The presence of these multiparticle
states at $p^{2}<0$ instead indicates that there might be instabilities, e.g, related to a spontaneous symmetry breaking~\cite{Sen:2002an}.} on the real axis, at $|\mathrm{Re}(z)|\geq1$, i.e., for $|\mathrm{q^{2}}|\geq m_{th}^{2}$.
The pole structure of the propagator~\eqref{eq:goodpropy} and the branch
cuts in the $q^{2}$- and the $q_{0}$-complex-energy planes are shown
in Fig.~\ref{fig:Poles-structure} for $\alpha=-0.4$. For $\alpha<0$,
the are no additional ghost-like degrees of freedom. In particular
there are no poles in the first and third quadrants of the complex
$q_{0}$-plane. 
Moreover, since the function $z^{-1}\mathrm{arctanh}(z)$ vanishes asymptotically, i.e. as $|z|\to\infty$, the full propagator $D(q^{2})$
defined above scales as $1/q^{2}$ in this limit. Accordingly, at variance of the case of exponential form factors \cite{Modesto:2017sdr}, in this case there
are no essential singularities at infinity, neither in Lorentzian
nor in Euclidean. There are thus no obstructions towards performing an analytic Wick rotation connecting the Euclidean and Lorentzian theories. Finally,
the absence of complex-conjugate degrees of freedom and ghost-resonances
with negative width (Merlin modes, \cite{Donoghue:2018lmc}) implies
that there cannot be any violation of causality, not even on microscopic
scales. 

\section{Conclusions}\label{sect5}

In this letter we discussed aspects of non-perturbative unitarity in QFT, in relation with the truncation-method typically employed to solve the FRG equations. 
The motivation of this work comes from the attempt to quantize gravity within the framework of QFT: while there are strong indications that gravity could be asymptotically safe, not much is known about the unitarity of the theory. 

Solving the FRG equations allows one to investigate the renormalizability of field theories and to compute their effective action. The effective action, in turn, can  be used to compute fully-dressed scattering amplitudes and to determine the complete spectrum of asymptotic states of the theory. On the one hand, the effective action provides an alternative and straightforward way to assess unitarity of QFTs, avoiding any perturbative expansions. On the other hand, solving the FRG equations exactly is still out of reach and approximations have to be employed. In practice, one has to resort to approximations or truncations, and the question arises what one can learn about unitarity from the graviton propagator obtained from a truncated expansion. 
We have formulated simple criteria for a ``good propagator'' in a consistent unitary theory. Many graviton propagators proposed in the literature violate at least one of those criteria. We therefore have provided a simple example for a good propagator, in order to demonstrate that there in principle is no obstruction to asymptotic safety of gravity from this side. The example shows that the analytic structure of a good propagator can be subtle. Knowing a consistent short distance theory many of these ``subtleties'' could find a natural explanation. Without such knowledge, approximations easily lead to a graviton propagator that apparently violates unitarity. In fact, finite truncations of the theory space naturally leads to the appearance of several poles in the fully-dressed propagator, and thus to an apparent violation of unitarity. While it is clear that some of these poles could be a truncation artifact, it is an interesting question how to understand whether these poles would also appear in the full (untruncated) theory. Understanding this point represents a first important step towards understanding the nature of the spin-2 ghost of Stelle gravity, which could indeed be a truncation artifact rather than a feature of a QFT of gravity. 

Using an artificially-truncated version of one-loop QED and Lee-Wick QED as working examples, we discovered that the truncation dependence of the propagator differs substantially between ghosts appearing only within truncations of the effective action (fake ghosts) and ghosts which also appear in the full, untruncated theory. While for the latter the residue remains always negative, in the former the residue is negative but its absolute value decreases with the truncation order and vanishes once all operators allowed by symmetry are included in the effective action. These fake ghosts disappear from the spectrum of asymptotic states of the theory. Our results lead us to the conjecture that, even not knowing the form of the quantum effective action, it might be possible to determine the nature of an apparent ghost-pole by tracing the behavior of the corresponding residue as function of the truncation order: if its residue is negative and stays negative for any value of the truncation order, then the pole corresponds to a genuine degree of freedom of the model and indicates a lack of unitarity. If instead the residue decreases with the truncation order and tends to zero when a sufficiently large number of terms is included in the action, it is likely that it corresponds to a fake ghost, i.e., a fictitious degree of freedom generated by the truncation of the theory space. It will be interesting to see if high order derivative expansion for quantum gravity can be used for an investigation in that direction. 


\begin{acknowledgments}
{The authors thank D. Anselmi, J. Donoghue and A. Eichhorn for insightful discussions, and A. Eichhorn for initial inputs on the development of this work. The research of AP is supported by the Alexander von Humboldt Foundation.}
\end{acknowledgments}

\bibliography{AleBib}

\end{document}